\documentclass[11pt,letterpaper]{article}

\usepackage[latin1]{inputenc}
\usepackage[dvips]{graphicx}
\usepackage{color}
\usepackage{amsfonts}

\usepackage[margin=1cm,font=small,labelfont=bf]{caption}

\newcommand{\qed}{\hfill$\Box$ \vspace{0.5 cm}}
\newenvironment{proof}{{\it Proof~: }}{\qed}

\newcommand{\point}{\textrm{\Huge{\hspace{-1pt}.}}}

\newtheorem{property}{Property}
\newtheorem{definition}{Definition}
\newtheorem{theorem}{Theorem}
\newtheorem{lemma}{Lemma}

\begin{document}

\begin{center}
{\Large{\bf Computing communities in large networks using random walks}}\\
\vspace{0.5cm}
{\Large Matthieu Latapy and Pascal Pons}\\ \vspace{2mm}
LIAFA -- CNRS and university Paris 7\\
2 place Jussieu, 75251 Paris Cedex 05, France\\
(latapy, pons)@liafa.jussieu.fr
\end{center}

\abstract{ 
{\footnotesize Dense subgraphs of sparse graphs (\emph{communities}), which appear in most real-world complex networks, play an important role in many contexts. Computing them however is generally expensive. We propose here a measure of similarities between vertices based on random walks which has several important advantages: it captures well the community structure in a network, it can be computed efficiently, it works at various scales, and it can be used in an agglomerative algorithm to compute efficiently the community structure of a network. We propose such an algorithm which runs in time $O(mn^2)$ and space $O(n^2)$ in the worst case, and in time $O(n^2\log n)$ and space $O(n^2)$ in most real-world cases ($n$ and $m$ are respectively the number of vertices and edges in the input graph). Experimental evaluation shows that our algorithm surpasses previously proposed ones concerning the quality of the obtained community structures and that it stands among the best ones concerning the running time. This is very promising because our algorithm can be improved in several ways, which we sketch at the end of the paper.}}\\

\section{Introduction}
Recent advances have brought out the importance of \emph{complex networks} in many different domains such as sociology (acquaintance networks, collaboration networks), biology (metabolic networks, gene networks) or computer science (Internet topology, Web graph, P2P networks). We refer to \cite{wasserman94socialnetwork,Strogatz:2001,albert:2002,Newman:2003,Dorogovtsev:2003} for reviews from different perspectives and for an exhaustive bibliography. The associated graphs are in general globally sparse but locally dense: there exist groups of vertices, called \emph{communities}, highly connected between them but with few links to other vertices. This kind of structure brings out much information about the network. For example, in a metabolic network the communities correspond to biological functions of the cell \cite{Ravasz:2002}. In the Web graph the communities correspond to topics of interest \cite{kleinberg:2001,Flake:2002}.

This notion of community is however difficult to define formally. Many definitions have been proposed in social networks studies \cite{wasserman94socialnetwork}, but they are too restrictive or cannot be computed efficiently. However, most recent approaches have reached a consensus, and consider that a partition $\mathcal{P} = \{C_1, \dots, C_k\}$ of the vertices of a graph $G = (V,E)$ ($\forall i, C_i \subseteq V$) represents a good community structure if the proportion of edges inside the $C_i$ (internal edges) is high compared to the proportion of edges between them (see for example the definitions given in \cite{Fortunato:2004}). Therefore, we will design an algorithm which finds communities satisfying this criterion.

We will consider throughout this paper an \emph{undirected graph} $G = (V,E)$ with $n = |V|$ vertices and $m = |E|$ edges. We impose that each vertex is linked to itself by a loop (we add these loops if necessary). We also suppose that $G$ is connected, the case where it is not being treated by considering the components as different graphs.

\subsection{Our approach and results}

Our approach is based on the following intuition, already pointed out in \cite{Gaume:2004}: small length random walks on a graph tend to get ``trapped'' into densely connected parts corresponding to communities. We therefore begin with a theoretical study of random walks on graphs. Using this, we define a measurement of the structural similarity between vertices and between communities, thus defining a distance. We relate this distance to existing spectral approaches of the problem. But our distance has an important advantage on these methods: it is efficiently computable, and can be used in a hierarchical clustering algorithm (merging iteratively the vertices into communities). One obtains this way a hierarchical community structure that may be represented as a tree structure called \emph{dendrogram} (an example is provided in Figure~\ref{figure:example}). We propose such an algorithm which finds a community structure in time $\mathcal{O}(mnH)$ where $H$ is the height of the corresponding dendrogram. The worst case is $\mathcal{O}(mn^2)$. But most real-world complex networks are sparse ($m = \mathcal{O}(n)$) and, as already noticed in \cite{Clauset_Newman:2004}, $H$ is generally small and tends to the most favourable case in which the dendrogram is balanced ($H = \mathcal{O}(\log n)$). In this case, the complexity is therefore $\mathcal{O}(n^2\log n)$. We finally evaluate the performance of our algorithm with different experiments which show that it surpasses previously proposed algorithms.

\subsection{Related work}

There exist many algorithms to find community structure in graphs. Most of them result from very recent works, but this topic is related to the classical problem of \emph{graph partitioning} that consists in splitting a graph into a given number of groups while minimizing the cost of the edge cut \cite{Fiedler:1973,Pothen:1990,Kernighan:1970}. However, these algorithms are not well suited to our case because they need the number of communities and their size as parameters. The recent interest in the domain has started with a new \emph{divisive} approach proposed by Girvan and Newman \cite{Girvan_Newman:2002, Newman_Girvan:2004}: the edges with the largest \emph{betweenness} (number of shortest paths passing through an edge) are removed one by one in order to split hierarchically the graph into communities. This algorithm runs in time $\mathcal{O}(m^2n)$. Similar algorithms were proposed by Radicchi \textit{et al} \cite{Radicchi_Filippo:2004} and by Fortunato \textit{et al} \cite{Fortunato:2004}. The first one uses a local quantity (the number of loops of a given length containing an edge) to choose the edges to remove and runs in time $\mathcal{O}(m^2)$. The second one uses a more complex notion of information centrality that gives better results but poor performances in $\mathcal{O}(m^3n)$.

\emph{Hierarchical clustering} is another classical approach introduced by sociologists for data analysis \cite{Aldenderfer:1984,Everitt:2001}. From a measurement of the similarity between vertices, an \emph{agglomerative} algorithm groups iteratively the vertices into communities (there exist different methods differing on the way of choosing the communities to merge at each step). We will use this approach in our algorithm and other agglomerative methods have also been recently introduced. Newman proposed in \cite{Newman:2004} a greedy algorithm that starts with $n$ communities corresponding to the vertices and merges communities in order to optimize a function called modularity which measures the quality of a partition. This algorithm runs in $\mathcal{O}(mn)$ and has recently been improved to a complexity $\mathcal{O}(mH\log n)$ (with our notations) \cite{Clauset_Newman:2004}. The algorithm of Donetti and Mu\~noz \cite{Donetti:2004} uses a hierarchical clustering method: they use the eigenvectors of the Laplacian matrix of the graph to measure the similarities between vertices. The complexity is determined by the computation of all the eigenvectors, in $\mathcal{O}(n^3)$ time for sparse matrices.

In the current situation, one can process graphs with up to a few hundreds of thousands vertices using the method in \cite{Clauset_Newman:2004}. All other algorithms have more limited performances (they generally cannot manage more than some thousands of vertices).

\section{Preliminaries on random walks}

The graph G is associated to its \emph{adjacency matrix} $A$: $A_{ij} = 1$ if vertices $i$ and $j$ are connected and $A_{ij} = 0$ otherwise. The degree $d(i) = \sum_j A_{ij}$ of the vertex $i$ is the number of its neighbors (including itself). To simplify the notations, we only consider unweighted graphs in this paper. It is however trivial to extend our results to weighted graphs ($A_{ij} \in \mathbb{R}^+$ instead of  $A_{ij} \in \{0,1\}$), which is an advantage of this approach.\\

Let us consider a discrete \emph{random walk process} (or diffusion process) on the graph $G$ (see \cite{Lovasz_random_walks,book_Aldous} for a complete presentation of the topic). At each time step a walker is on a vertex and moves to a vertex chosen randomly and uniformly among its neighbors. The sequence of visited vertices is a \emph{Markov chain}, the states of which are the vertices of the graph. At each step, the transition probability from vertex $i$ to vertex $j$ is $P_{ij} = \frac{A_{ij}}{d(i)}$. This defines the \emph{transition matrix} $P$ of the random walk. We can also write $P = D^{-1}A$ where $D$ is the diagonal matrix of the degrees ($\forall i, D_{ii} = d(i)$ and $D_{ij} = 0$ for $i \neq j$).

The process is driven by the powers of the matrix $P$: the probability of going from $i$ to $j$ through a random walk of length $t$ is $(P^t)_{ij}$. In the following, we will denote this probability by $P_{ij}^t$. It satisfies two general properties of the random walk process (see proofs in Appendix A) which we will use in the sequel:

\begin{property} \label{limit_proba}
When the length $t$ of a random walk starting at vertex $i$ tends towards infinity, the probability of being on a vertex $j$ only depends on the degree of vertex $j$ (and not on the starting vertex $i$):
$$\forall i, \lim_{t \rightarrow +\infty} P_{ij}^t = \frac{d(j)}{\sum_{k}d(k)}$$
\end{property}

\begin{property} \label{symetry_proba}
The probabilities of going from $i$ to $j$ and from $j$ to $i$ through a random walk of a fixed length $t$ have a ratio that only depends on the degrees $d(i)$ and $d(j)$:
$$
\forall i, \forall j, d(i) P_{ij}^t = d(j) P_{ji}^t
$$
\end{property}

\section{Comparing vertices using short random walks}

In order to group the vertices into communities, we will now introduce a distance $r$ between the vertices that reflects the community structure of the graph. This distance must be large if the two vertices are in different communities, and on the contrary if they are in the same community it must be small. It will be computed from the information given by random walks in the graph.

Let us consider random walks on $G$ of a given length $t$. We will use the information given by all the probabilities $P_{ij}^t$ to go from $i$ to $j$ in $t$ steps. The length $t$ of the random walks must be sufficiently long to gather enough information about the topology of the graph. However $t$ must not be too long, to avoid the effect predicted by Property~\ref{limit_proba}; the probabilities would only depend on the degree of the vertices. Each probability $P_{ij}^t$ gives some information about the two vertices $i$ and $j$, but Property~\ref{symetry_proba} says that $P_{ij}^t$ and $P_{ji}^t$ encode exactly the same information. Finally, the information about vertex $i$ encoded in $P^t$ resides in the $n$ probabilities $(P_{ik}^t)_{1 \leq k \leq n}$, which is nothing but the $i^{\textrm{\tiny{th}}}$ row of the matrix $P^t$, denoted by $P_{i\point}^t$. To compare two vertices $i$ and $j$ using these data, we must notice that:
\begin{itemize}
\item If two vertices $i$ and $j$ are in the same community, the probability $P_{ij}^t$ will surely be high. But the fact that $P_{ij}^t$ is high does not necessarily imply that $i$ and $j$ are in the same community.
\item The probability $P_{ij}^t$ is influenced by the degree $d(j)$ because the walker has higher probability to go to high degree vertices.
\item Two vertices of a same community tend to ``see'' all the other vertices in the same way. Thus if $i$ and $j$ are in the same community, we will probably have $\forall k, P_{ik}^t \simeq P_{jk}^t$.\\
\end{itemize}
We can now give the definition of our distance between vertices, which takes into account all previous remarks:
\begin{definition} Let $i$ and $j$ be two vertices in the graph:
\begin{equation} \label{definition_distance}
r_{ij} = \sqrt{\sum_{k = 1}^n\frac{(P_{ik}^t - P_{jk}^t)^2}{d(k)}} = \Big\|D^{-\frac{1}{2}}P_{i\point}^t - D^{-\frac{1}{2}}P_{j\point}^t \Big\| 
\end{equation}
where $\|.\|$ is the Euclidean norm of  $\mathbb{R}^n$.
\end{definition}
One can notice that this distance can also be seen as the $L^2$ distance \cite{book_Aldous} between the two probability distributions $P_{i\point}^t$ and $P_{j\point}^t$. Notice also that the distance depends on $t$ and may be denoted $r_{ij}(t)$. We will however consider it as implicit to simplify the notations.

\begin{theorem} \label{th_distance}
The distance $r$ is related to the spectral properties of the matrix $P$ by:
$$
r_{ij}^2 = \sum_{\alpha = 2}^n \lambda_{\alpha}^{2t} (v_{\alpha}(i) - v_{\alpha}(j))^2
$$
where $(\lambda_{\alpha})_{1 \leq \alpha \leq n}$ and $(v_{\alpha})_{1 \leq \alpha \leq n}$ are respectively the eigenvalues and right eigenvectors of the matrix $P$.
\end{theorem}
In order to prove this theorem, we need the following technical lemma:
\begin{lemma} \label{lemma}
The eigenvalues of the matrix $P$ are real and satisfy:
$$
1 = \lambda_1 > \lambda_2 \geq \dots \geq \lambda_n > -1
$$
Moreover, there exists an orthonormal family of vectors $(s_{\alpha})_{1 \leq \alpha \leq n}$ such that each vector $v_{\alpha} = D^{-\frac{1}{2}} s_{\alpha}$ and $u_{\alpha} = D^{\frac{1}{2}} s_{\alpha}$ are respectively a right and a left eigenvector associated to the eigenvalue $\lambda_{\alpha}$:
$$
\forall \alpha, P v_{\alpha} = \lambda_{\alpha} v_{\alpha} \textrm{ and } P^T u_{\alpha} = \lambda_{\alpha} u_{\alpha}
$$
$$
\forall \alpha, \forall \beta, v_{\alpha}^T u_{\beta} = \delta_{\alpha\beta}
$$
\end{lemma}
\begin{proof}The matrix $P$ has the same eigenvalues as its similar matrix $S = D^{\frac{1}{2}}PD^{-\frac{1}{2}} = D^{-\frac{1}{2}}AD^{-\frac{1}{2}}$. The matrix $S$ is real and symmetric, so its eigenvalues $\lambda_{\alpha}$ are real. $P$ is a stochastic matrix ($\sum_{j=1}^n P_{ij} = 1$), so its largest eigenvalue is $\lambda_1 = 1$. The graph $G$ is connected and primitive (the $\gcd$ of the cycle lengths of $G$ is $1$, due to the loops on each vertex), therefore we can apply the Perron-Frobenius theorem which implies that $P$ has a unique dominant eigenvalue. Therefore we have: $|\lambda_{\alpha}| < 1$ for $2 \leq \alpha \leq n$.

The symmetry of $S$ implies that there also exists an orthonornal family $s_{\alpha}$ of eigenvectors of $S$ satisfying $\forall \alpha, \forall \beta, s_{\alpha}^Ts_{\beta} = \delta_{\alpha\beta}$ (where $\delta_{\alpha\beta} = 1$ if $\alpha = \beta$ and $0$ otherwise). We then directly obtain that the vectors $v_{\alpha} = D^{-\frac{1}{2}} s_{\alpha}$ and $u_{\alpha} = D^{\frac{1}{2}} s_{\alpha}$ are respectively a right and a left eigenvector of $P$ satisfying $u_{\alpha}^Tv_{\beta} = \delta_{\alpha\beta}$.
\end{proof}\\
We can now prove Theorem~\ref{th_distance}:\\
\begin{proof}Lemma \ref{lemma} makes it possible to write a spectral decomposition of the matrix $P$: 
$$
P = \sum_{\alpha = 1}^n \lambda_{\alpha}v_{\alpha}u_{\alpha}^T \textrm{, and } P^t = \sum_{\alpha = 1}^n \lambda_{\alpha}^tv_{\alpha}u_{\alpha}^T \textrm{, and so } P_{ij}^t = \sum_{\alpha = 1}^n \lambda_{\alpha}^tv_{\alpha}(i)u_{\alpha}(j)
$$
Now we obtain the expression of the probability vector $P_{i\point}^t$:
$$
P_{i\point}^t = \sum_{\alpha = 1}^n \lambda_{\alpha}^t v_{\alpha}(i) u_{\alpha} = D^{\frac{1}{2}} \sum_{\alpha = 1}^n \lambda_{\alpha}^t v_{\alpha}(i) s_{\alpha}
$$
We put this formula into the second definition of $r_{ij}$ given in Equation (\ref{definition_distance}). Then we use the Pythagorean theorem with the orthonormal family of vectors $(s_{\alpha})_{1 \leq \alpha \leq n}$, and we remember that the vector $v_1$ is constant to remove the case $\alpha = 1$ in the sum. Finally we have:
$$
r_{ij}^2 = \bigg\|\sum_{\alpha = 1}^n \lambda_{\alpha}^t (v_{\alpha}(i) - v_{\alpha}(j)) s_{\alpha} \bigg\|^2 = \sum_{\alpha = 2}^n \lambda_{\alpha}^{2t} (v_{\alpha}(i) - v_{\alpha}(j))^2
$$
\end{proof}

This theorem relates random walks on graphs to the many current works that study spectral properties of graphs. For example, \cite{Simonsen:2004} notices that the modular structure of a graph is expressed in the eigenvectors of $P$ (other than $v_1$) that corresponds to the largest positive eigenvalues. If two vertices $i$ and $j$ belong to a same community then the coordinates $v_{\alpha}(i)$ and $v_{\alpha}(j)$ are similar in all these eigenvectors. Moreover, \cite{Schulman:2001,Gaveau:1999} show in a more general case that when an eigenvalue $\lambda_{\alpha}$ tends to $1$, the coordinates of the associated eigenvector $v_{\alpha}$ are constant in the subsets of vertices that correspond to communities. A distance similar to ours (but that cannot be computed directly with random walks) is also introduced: $d_{t}^2(i,j) = \sum_{\alpha = 2}^{n} \frac{(v_{\alpha}(i) - v_{\alpha}(j))^2}{1 - |\lambda_{\alpha}|^t}$. Finally, \cite{Donetti:2004} uses the same spectral approach applied to the Laplacian matrix of the graph $L = D - A$.

All these studies show that the spectral approach takes an important part in the search for community structure in graphs. However all these approaches have the same drawback: the eigenvectors need to be explicitly computed (in time $\mathcal{O}(n^3)$ for a sparse matrix). This computation rapidly becomes untractable in practice when the size of the graph exceeds some thousands of vertices. Our approach is based on the same foundation but has the advantage of avoiding the expensive computation of the eigenvectors: it only needs to compute the probabilities $P_{ij}^t$, which can be done efficiently as shown in the following theorem.

\begin{theorem}
All the probabilities $P_{ij}^t$ can be computed in time $\mathcal{O}(tnm)$ and space $\mathcal{O}(n^2)$. Once these probabilities computed, each distance $r_{ij}$ can be computed in time $\mathcal{O}(n)$. For given $i$ and $j$, one can also compute directly $r_{ij}$ in time $\mathcal{O}(tm)$ and space $\mathcal{O}(n)$.
\end{theorem}
\begin{proof}To compute the vector $P_{i\point}^t$, we multiply $t$ times the vector $P_{i\point}^0$ ($\forall k, P_{i\point}^0(k) = \delta_{ik}$) by the matrix $P$. This direct method is advantageous in our case because the matrix $P$ is generally sparse (for real-world complex networks) therefore each product is processed in time $\mathcal{O}(m)$. The initialization of $P_{i\point}^0$ is done in $\mathcal{O}(n)$ and thus each of the $n$ vectors $P_{i\point}^t$ is computed in time $\mathcal{O}(n + tm) = \mathcal{O}(tm)$. Once we have the two vectors $P_{i\point}^t$ and $P_{j\point}^t$, we can compute $r_{ij}$ in $\mathcal{O}(n)$ using Equation (\ref{definition_distance}). We can compute and keep in memory all the probability vectors in time $\mathcal{O}(tnm)$ or compute directly $r_{ij}$ by evaluating the two vectors $P_{i\point}^t$ and $P_{j\point}^t$ in time $\mathcal{O}(tm)$.
\end{proof}

Now we generalize our distance between vertices to a distance between communities in a straightforward way. Let us consider random walks that start from a community: the starting vertex is chosen randomly and uniformly among the vertices of the community. We define the probability $P_{Cj}^t$ to go from community $C$ to vertex $j$ in $t$ steps:
$$
P_{Cj}^t = \frac{1}{|C|}\sum_{i \in C} P_{ij}^t
$$
This defines a probability vector $P_{C\point}^t$ that allows us to generalize our distance:
\begin{definition}
Let $C_1, C_2 \subset V$ be two communities. We define the distance $r_{C_1C_2}$ between these two communities by:
$$
r_{C_1C_2} = \Big\| D^{-\frac{1}{2}}P_{C_1\point}^t - D^{-\frac{1}{2}}P_{C_2\point}^t \Big\| = \sqrt{\sum_{k = 1}^n\frac{(P_{C_1k}^t - P_{C_2k}^t)^2}{d(k)}}
$$
\end{definition}
This definition is consistent with the previous one: $r_{ij} = r_{\{i\}\{j\}}$ and we can also define the distance between a vertex $i$ and a community $C$: $r_{iC} = r_{\{i\}C}$. Given the probability vectors $P_{C_1\point}^t$ and $P_{C_2\point}^t$, the distance $r_{C_1C_2}$ is also computed in time $\mathcal{O}(n)$.

\section{The algorithm}

In the previous section, we have proposed a distance between vertices (and between sets of vertices) which captures structural similarities between them. The problem of finding communities is now a clustering problem. We will use here an efficient hierarchical clustering algorithm that allows us to find community structures at different scales. We present an agglomerative approach based on Ward's method \cite{Ward:1963} that is well adapted to our distance and gives very good results while reducing the number of distance computations in order to be able to process large graphs.

\begin{figure}[ht]
\begin{center}
\includegraphics{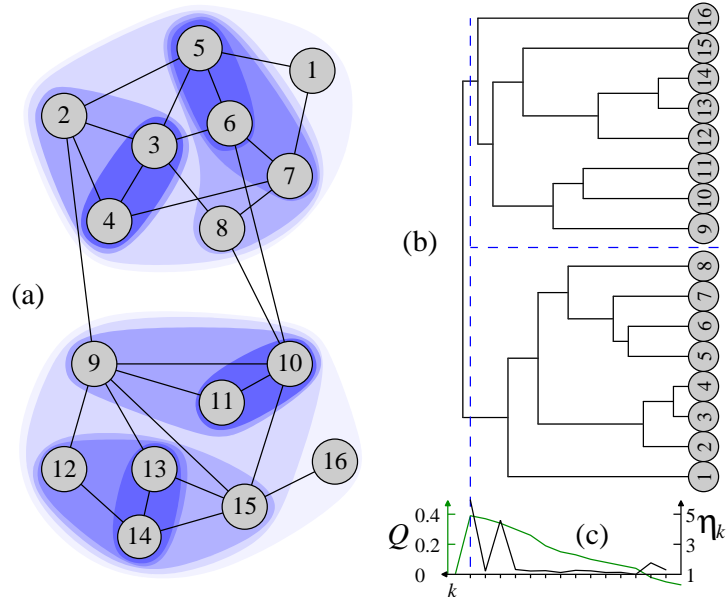}
\caption{(a) An example of community structure found by our algorithm using random walks of length $t = 3$. (b) The stages of the algorithm encoded as a tree called \emph{dendrogram}. The maximum of $\eta_k$ and $Q$, plotted in (c), show that the best partition consists in two communities.}
\label{figure:example}
\end{center}
\end{figure}

We start from a partition $\mathcal{P}_1 = \{\{v\}, v \in V\}$ of the graph into $n$ communities reduced to a single vertex. We first compute the distances between all adjacent vertices. Then this partition evolves by repeating the following operations. At each step $k$:
\begin{itemize}
\item Choose two communities $C_1$ and $C_2$ in $\mathcal{P}_k$ on a criterion based on the distance between the communities that we detail later.
\item Merge these two communities into a new community $C_3 = C_1 \cup C_2$ and create the new partition: $\mathcal{P}_{k+1} = (\mathcal{P}_k \setminus \{C_1, C_2\}) \cup \{C_3\}$.
\item Update the distances between communities (we will see later that we actually only do this for {\em adjacent} communities).
\end{itemize}
After $n-1$ steps, the algorithm finishes and we obtain $\mathcal{P}_n = \{V\}$. Each step defines a partition $\mathcal{P}_k$ of the graph into communities, which gives a hierarchical structure of communities called dendrogram (see Figure~\ref{figure:example}(b)). This structure is a tree in which the leaves correspond to the vertices and each internal node is associated to a merging of communities in the algorithm: it corresponds to a community composed of the union of the communities corresponding to its children. 

The key points in this algorithm are the way we choose the communities to merge, and the fact that the distances can be updated efficiently. We will also need to evaluate the quality of a partition in order to choose one of the $\mathcal{P}_k$ as the result of our algorithm. We will detail these points below, and explain how they can be managed to give an efficient algorithm.

\paragraph{Choosing the communities to merge.} This choice plays a central role for the quality of the community structure created. In order to reduce the complexity, we will only merge {\em adjacent} communities (having at least an edge between them). This reasonable heuristic (already used in \cite{Newman:2004} and \cite{Donetti:2004}) limits to $m$ the number of possible mergings at each stage. Moreover it ensures that each community is connected.

We choose the two communities to merge according to Ward's method. At each step $k$, we merge the two communities that minimize the mean $\sigma_k$ of the squared distances between each vertex and its community. 
$$
\sigma_k = \frac{1}{n} \sum_{C \in \mathcal{P}_k} \sum_{i\in C} r_{iC}^2
$$
This approach is a greedy algorithm that tries to solve the problem of maximizing $\sigma_k$ for each $k$. But this problem is known to be NP-hard: even for a given $k$, maximizing $\sigma_k$ is the NP-hard ``K-Median clustering problem'' \cite{Fernandez_de_la_Vega:2003, Drineas:2004} for $K = (n-k)$ clusters. The existing approximation algorithms \cite{Fernandez_de_la_Vega:2003, Drineas:2004} are exponential with the number of clusters to find and unsuitable for our purpose. 
So for each pair of adjacent communities $\{C_1,C_2\}$, we compute the variation $\Delta\sigma(C_1,C_2)$ of $\sigma$ if we would merge $C_1$ and $C_2$ into a new community $C_3 = C_1 \cup C_2$. This quantity only depends on the vertices of $C_1$ and $C_2$, and not on the other communities or on the step $k$ of the algorithm:
\begin{equation} \label{def_delta_sigma}
\Delta\sigma(C_1,C_2) = \frac{1}{n}\Big(\sum_{i\in C_3} r_{iC_3}^2 - \sum_{i\in C_1} r_{iC_1}^2 - \sum_{i\in C_2} r_{iC_2}^2\Big)
\end{equation}
Finally, we merge the two communities that give the lowest value of $\Delta\sigma$. 

\paragraph{Computing $\Delta\sigma$ and updating the distances.} The important point here is to notice that these quantities can be efficiently computed thanks to the fact that our distance is a Euclidean distance, which makes it possible to obtain the two following classical results \cite{Jambu} (proofs in Appendix A):
\begin{theorem} \label{th_delta_sigma_1} The increase of $\sigma$ after the merging of two communities $C_1$ and $C_2$ is directly related to the distance $r_{C_1C_2}$ by:
$$
\Delta\sigma(C_1,C_2) = \frac{1}{n}\frac{|C_1||C_2|}{|C_1| + |C_2|} r_{C_1C_2}^2
$$
\end{theorem}

This theorem shows that we only need to update the distances between communities to get the values of $\Delta\sigma$: if we know the two vectors $P_{C_1\point}$ and $P_{C_2\point}$, the computation of $\Delta\sigma(C_1,C_2)$ is possible in $\mathcal{O}(n)$. Moreover, the next theorem shows that if we already know the three values $\Delta\sigma(C_1,C_2)$, $\Delta\sigma(C_1,C)$ and $\Delta\sigma(C_2,C)$, then we can compute $\Delta\sigma(C_1 \cup C_2, C)$ in constant time.
\begin{theorem}[Lance-Williams-Jambu formula] \label{th_delta_sigma_2} If $C_1$ and $C_2$ are merged into $C_3 = C_1 \cup C_2$ then for any other community $C$:
\begin{equation} \label{eq_th_delta_sigma_2}
\Delta\sigma(C_3, C) = \frac{(|C_1| + |C|) \Delta\sigma(C_1,C) + (|C_2| + |C|) \Delta\sigma(C_2,C) - |C| \Delta\sigma(C_1,C_2)}{|C_1|+|C_2|+|C|}
\end{equation}
\end{theorem}

Since we only merge adjacent communities, we only need to update the values of $\Delta\sigma$ between adjacent communities (there are at most $m$ values). These values are stored in a balanced tree in which we can add, remove or get the minimum in $\mathcal{O}(\log m)$. Each computation of a value of $\Delta\sigma$ can be done in time $\mathcal{O}(n)$ with Theorem~\ref{th_delta_sigma_1} or in constant time when Theorem~\ref{th_delta_sigma_2} can be applied.

\paragraph{Evaluating the quality of a partition.} The algorithm induces a sequence $(\mathcal{P}_k)_{1 \leq k \leq n}$ of partitions into communities. We now want to know which partitions in this sequence are good representations of communities. The most common way is to use a statistical parameter such as the modularity $Q$ introduced in \cite{Newman_Girvan:2004,Newman:2004}. This quantity (between $-1$ and $1$) is well suited to find the best partition but not to find several ones (corresponding to other scales in the hierarchical structure, see Appendix B). Here we provide another criterion that helps in finding different scales of communities. When we merge two very different communities (with respect to the distance $r$), the value $\Delta\sigma_k = \sigma_{k+1} - \sigma_{k}$ at this step is large. Conversely, if $\Delta\sigma_k$ is large then the communities at step $k-1$ are surely relevant. To detect this, we introduce the increase ratio $\eta_k$:
$$
\eta_k = \frac{\Delta\sigma_k}{\Delta\sigma_{k-1}} = \frac{\sigma_{k+1} - \sigma_k}{\sigma_k - \sigma_{k-1}}
$$
We then assume that the best partitions $\mathcal{P}_k$ are those associated with the largest values of $\eta_k$. Depending on the context in which our algorithm is used, one may take only the best partition (the one for which $\eta_k$ is maximal) or choose among the best ones using another criterion (like the size of the communities, for instance). This is an important advantage of our method, which gives different scales in the community structure, as illustrated in Appendix B.

\vspace{-3mm}
\paragraph{Complexity.} 
First, the initialization of the probability vectors is done in $\mathcal{O}(mnt)$. Then, at each step $k$ of the algorithm, we keep in memory the vectors $P_{C\point}^t$ corresponding to the current communities (the ones in the current partition). But for the communities that are not in $\mathcal{P}_k$  (because they have been merged with another community before) we only keep the information saying in which community it has been merged. We keep enough information to construct the dendogram and have access to the composition of any community with a few more computation.

\noindent
When we merge two communities $C_1$ and $C_2$ we perform the following operations:
\begin{itemize}
\item Compute $P_{(C_1\cup C_2)\point}^t = \frac{ |C_1| P_{C_1\point}^t +  |C_2| P_{C_2\point}^t}{ |C_1|+|C_2|}$ and remove $ P_{C_1\point}^t$ and $P_{C_2\point}^t$.
\item Update the values of $\Delta\sigma$ concerning $C_1$ and $C_2$ using Theorem~\ref{th_delta_sigma_2} if possible, or otherwise using Theorem~\ref{th_delta_sigma_1}.
\end{itemize}
The first operation can be done in $\mathcal{O}(n)$, and therefore does not play a significant role in the overall complexity of the algorithm. The dominating factor in the complexity of the algorithm is the number of distances $r$ computed (each one in $\mathcal{O}(n)$). We prove an upper bound of this number that depends on the height of the dendrogram. We denote by $h(C)$ the height of a community $C$ and by $H$ the height of the whole tree ($H = h(V)$).

\begin{theorem}
An upper bound of the number of distances computed by our algorithm is $2mH$. Therefore its global time complexity is $\mathcal{O}(mn(H+t))$.
\end{theorem}
\begin{proof}Let $M$ be the number of computations of $\Delta\sigma$. $M$ is equal to $m$ (initialization of the first $\Delta\sigma$) plus the sum over all steps $k$ of the number of neighbors of the new community created at step $k$ (when we merge two communities, we need to update one value of $\Delta\sigma$ per neighbor). For each height $1 \leq h \leq H$, the communities with the same height $h$ are pairwise disjoint, and the sum of their number of neighbor communities is less than $2m$ (each edge can at most define two neighborhood relations). The sum over all heights finally gives $M \leq 2Hm$. Each of these $M$ computations needs at most one computation of $r$ in time $\mathcal{O}(n)$ (Theorem~\ref{th_delta_sigma_1}). Therefore, with the initialization, the global complexity is $\mathcal{O}(mn(H+t))$.
\end{proof}

In practice, a small $t$ must be chosen ($t = \mathcal{O}(\log n)$, because if it is not the case the random walks converge to the limit distribution at exponential speed) and thus the global complexity is $\mathcal{O}(mnH)$. The worst case is $H = n-1$, which occurs when the vertices are merged one by one to a large community. This happens in the ``star'' graph, where a central vertex is linked to the $n-1$ others. However Ward's algorithm is known to produce small communities of similar sizes. This tends to get closer to the favorable case in which the community structure is a balanced tree and its height is $H = \mathcal{O}(\log n)$.

\section{Experimental evaluation of the algorithm}

Evaluating a community detection algorithm is difficult because one needs some test graphs whose community structure is already known. A classical approach, which we will follow here, is to use randomly generated graphs with communities defined as follows: one constructs a graph with $n$ vertices and $c \geq 1$ disjoint communities of $\frac{n}{c}$ vertices. An internal and an external density of edges $p_{in}$ and $p_{out}$ are given. Each possible edge inside a community is drawn with probability $p_{in}$ and each possible edge between two communities is drawn with probability $p_{out}$. These two probabilities define an expected average in-degree $z_{in} = p_{in}(\frac{n}{c} - 1)$ and an expected average out-degree $z_{out} = p_{out}\frac{n(c-1)}{c}$. 

\begin{figure}[ht]
\begin{center}
\includegraphics{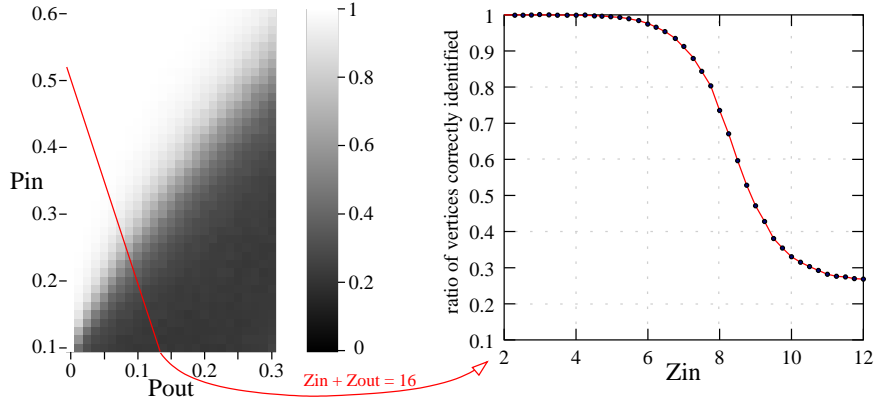}
\caption{Performances of our algorithm on graphs with $n = 128$ vertices and $c = 4$ communities using random walks of length $t=3$. Left: ratio of vertices correctly identified as a function of the edges density $p_{in}$ and $p_{out}$ (black stands for $0\%$ and white for $100\%$ ). Right: detail of this plot for a constant average degree $z_{in} + z_{out} = 16$.}
\label{figure:influence_density}
\end{center}
\end{figure}

In order to evaluate the performance of our algorithm we will evaluate the ratio of vertices correctly identified by the algorithm. This ratio has been used (without formal definition) in \cite{Fortunato:2004,Girvan_Newman:2002,Newman_Girvan:2004,Newman:2004,Donetti:2004}. Here we define it according to the following identification procedure: we want to identify the $c$ known communities $(C_{i})_{1 \leq i \leq c}$ to $\overline{c}$ communities $(\overline{C}_{j})_{1 \leq j \leq \overline{c}}$ found by the algorithm. We identify each $C_i$ to the community $\overline{C}_{\gamma(i)}$ such that $|C_i \cap \overline{C}_{\gamma(i)}|$ is maximal. If there are $l > 1$ communities $C_{i_1}, \dots C_{i_l}$ identified to the same community $\overline{C_j}$ ($\gamma(i_1) = \dots = \gamma(i_l) = j$), then we only keep the identification of the community $C_{i_k}$ which maximizes $|C_{i_k} \cap \overline{C}_j|$. The other communities $C_{i_{k'}}$ are no more identified to any community. A vertex is then correctly identified if it belongs to the community found by the algorithm is identified to its actual community.

In order to compare our algorithm to the other known algorithms, we first study the influence of the densities $p_{in}$ and $p_{out}$ on the same graphs as the ones used in \cite{Fortunato:2004,Girvan_Newman:2002,Newman_Girvan:2004,Newman:2004,Donetti:2004}. They considered graphs of $n = 128$ and $c = 4$ communities but different densities $p_{in}$ and $p_{out}$. The results are plotted in Figure~\ref{figure:influence_density}. It indicates that our algorithm has perfect results when the graph has a clear community structure (\textit{i.e.} when $p_{in}$ is high and $p_{out}$ is low). When $p_{out}$ is high and $p_{in}$ is low, the graph does not really have a community structure, which explains why our algorithm does not find it. In intermediate cases, our algorithm has better performances than previously proposed algorithms, which generally have only been tested in the case $z_{in} + z_{out} = 16$.

\medskip

\begin{figure}[ht]
\begin{center}
\hspace*{0.3in}
\begin{tabular}{cc}
\begin{minipage}{3.6in}
\begin{center}
{\small
\begin{tabular}{|c|c|c|}
\hline
nb vertices & ratio of vertices correctly identified & time\\
\hline
100 & 99\% & 0.05s\\
\hline
300 & 93\% & 0.25s\\
\hline
1,000 & 90\% & 2.6s\\
\hline
3,000 & 73\% & 21s\\
\hline
10,000 & 71\% & 11min\,\footnote{This case has been slowed by the lack of memory on the $512$~MB RAM machine on which the experiments were run.}\\
\hline
\end{tabular}
}
\end{center}
\end{minipage}
&
\begin{minipage}{2.5in}
\begin{center}
\includegraphics{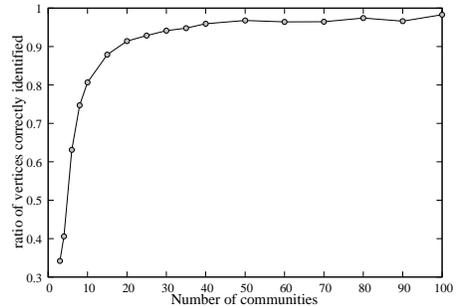}
\end{center}
\end{minipage}
\end{tabular}

\caption{Left: Performances of our algorithm on graphs with $c=10$ communities, $z_{out} = 8$ and $6 \leq z_{in} \leq 10$ for various sizes of graphs. Right: ratio of correctly identified vertices on the same kind of graphs with $n=5000$ vertices when the number of communities varies.} 
\label{experiments_large_graphs:example}
\end{center}
\end{figure}

In order to deepen the empirical study of the performances of our algorithm, we tested it on various situations. In all of them, the performances were very good, the only cases where our algorithm fails beeing the extreme cases in which no clear community structure exists. To illustrate this we detail two of these experiments below (and two other in Appendix B).
Let us consider graphs of different sizes from $n = 100$ to $n = 10,000$ with $c = 10$ communities. The external density of edges is chosen in order to have a mean out-degree $z_{out} = 8$. The internal density $p_{in}(C)$ of each community $C$ is randomly and uniformly chosen from an interval $[p_{min}..p_{max}]$ such that its mean internal degree satisfies $6 \leq z_{in}(C) \leq 10$. The results for $t = 5$ (Figure~\ref{experiments_large_graphs:example}, left) show that our algorithm has good performances on large graphs even with some heterogeneity in the communities.

We also studied the influence of the number of communities. We consider the same kind of graphs as above with $n = 5000$ vertices and with different number $c$ of communities. The results are plotted in Figure~\ref{experiments_large_graphs:example}(right). We chose to keep the same global density of graphs (the expected average degree is always $16$) and increase the number of communities which implies a decrease in their size and their internal density of edges. The experiment shows that, even if the overall number of expected external edges is equal to the one of internal edges, our algorithm easily detects the communities with a sufficient internal density.

\section{Improvements} \label{improvements}
 
Several improvements of our algorithm are possible and we investigated some. We do not present them in detail in this abstract, but rapidly outline two of the most interesting ones.

One may first replace the computation of the probabilities $P_{i\point}^t$ by approximation obtained by running a given number $K$ of random walks from vertex $i$. The precision of the estimation is $\mathcal{O}(\frac{1}{\sqrt K})$ and a good evaluation is typically obtained with $K = 1000$. This approach is interesting for very large graphs and allows us to estimate each vector $P_{i\point}^t$ in $\mathcal{O}(K(t+\log K))$.

Another improvement concerns the discrete nature of random walks. As already noticed, the best length $t$ of the random walks is generally small and its choice is difficult. The discrete time is restrictive and may not give enough freedom. One may then replace the discrete Markov chain by its continuous version, and obtain new probabilities $\widetilde{P}_{ij}^t$ which expression is $\forall t \in \mathbb{R}, \widetilde{P}_{ij}^t = \big(e^{t(P - Id)}\big)_{ij}$. Each probability vector $\widetilde{P}_{i\point}^t$ can be efficiently computed in $\mathcal{O}(rm)$ with an error $e^{-t}\sum_{k=r+1}^{+\infty}\frac{t^k}{k!}$ by truncating the exponential series at a given range $r$. This improvement makes it possible to choose non integer values for $t$.

\section{Conclusion and further work}

We proposed a new distance between the vertices that quantify their structural similarities using random walks. This distance has several advantages: it captures much information on the community structure, it is well suited for approximation, and it can be used in an efficient hierarchical agglomerative algorithm that detects communities in a network at different scales. We designed such an algorithm which works in time $\mathcal{O}(mn^2)$. In practice, real-world complex networks are sparse ($m = \mathcal{O}(n)$) and the height of the dendrogram is generally small ($H = \mathcal{O}(\log n)$); in this case the algorithm runs in $\mathcal{O}(n^2\log n)$. This complexity may be reduced with the improvements sketched in Section~\ref{improvements}.

Most previous methods were unable to manage networks with more than approximately $10,000$ vertices, except the one in \cite{Clauset_Newman:2004} which goes up to several hundreds of thousands. We ran our algorithm on networks of up to $100,000$ vertices, and experiments show that the obtained quality is better than the one obtained in \cite{Clauset_Newman:2004}, and  actually better than the one
obtained by all previous ones. Several possible improvements have been pointed out which will improve the performances of our algorithm. Moreover, our method is well suited for detectiong communities at various sacles (see Appendix B). We therefore think that it may be considered as a significant step in the area.

Choosing an appropriate length $t$ of the random walks is however still a problem and we have work in progress in this direction. More experiments on real-world complex networks also still have to be performed (see Appendix B), as well as direct comparison with algorithms proposed and implemented by other authors\,\footnote{They actually do not provide implementations of their algorithms, in general, which makes comparisons quite difficult. We make a step in this direction by providing an implementation of our method~\cite{url_prog}.}. Our approach may also be relevant for the computation of \emph{overlapping} communities (which often occurs in real-world cases and is not considered by any algorithm until now), which we consider as a promising direction for further work. Finally, we pointed out that the method is directly usable for {\em weighted} networks. For directed ones (like the important case of the Web graph), on the contrary, the proofs we provided are not valid anymore, and random walks behave significantly differently. Therefore, we also consider the directed case as an interesting direction.

\subsubsection*{Acknowledgments}

We warmly thank Ramesh Govindan and Damien Magoni for providing useful data. 
We also thank Annick Lesne and L. S. Shulman for useful conversation and Cl\'emence Magnien for helpful comments on preliminary versions.
This work has been supported in part by the PERSI ({\em Programme d'\'Etude des R\'eseaux Sociaux de l'Internet}) project and by the
GAP ({\em Graphs, Algorithms and Probabilities}) project.

\bibliography{biblio}

\begin{thebibliography}{10}

\bibitem{albert:2002}
R.~Albert and A.-L. Barab\'asi.
\newblock Statistical mechanics of complex networks.
\newblock {\em Reviews of Modern Physics}, 74(1):47, 2002.

\bibitem{Aldenderfer:1984}
M.~S. Aldenderfer and R.~K. Blashfield.
\newblock {\em Cluster Analysis}.
\newblock Number 07-044 in Sage University Paper Series on Quantitative
  Applications in the Social Sciences. Sage, Beverly Hills, 1984.

\bibitem{book_Aldous}
D.~Aldous and J.~A. Fill.
\newblock {\em Reversible Markov Chains and Random Walks on Graphs}, chapter~2.
\newblock Forthcoming book,
  http://www.stat.berkeley.edu/users/aldous/RWG/book.html.

\bibitem{Gaume:2004}
Gaume B.
\newblock Balades aléatoires dans les petits mondes lexicaux.
\newblock {\em I3 Information Interaction Intelligence}, (to appear).

\bibitem{Clauset_Newman:2004}
A.~Clauset, M.~E.~J. Newman, and C.~Moore.
\newblock Finding community structure in very large networks.
\newblock {\em arXiv:cond-mat/0408187}, 2004.

\bibitem{Donetti:2004}
L.~Donetti and M.~A. Mu\~noz.
\newblock Detecting network communities: a new systematic and efficient
  algorithm.
\newblock {\em arXiv:cond-mat/0404652}, 2004.

\bibitem{Dorogovtsev:2003}
S.N. Dorogovtsev and J.F.F. Mendes.
\newblock {\em Evolution of Networks: From Biological Nets to the Internet and
  WWW}.
\newblock Oxford University Press, Oxford, 2003.

\bibitem{Drineas:2004}
P.~Drineas, A.~Frieze, R.~Kannan, S.~Vempala, and V.~Vinay.
\newblock Clustering large graphs via the singular value decomposition.
\newblock {\em Machine Learning}, 56(1-3):9--33, 2004.

\bibitem{Everitt:2001}
B.~S. Everitt, S.~Landau, and M.~Leese.
\newblock {\em Cluster Analysis}.
\newblock Hodder Arnold, London, $4^{th}$ edition, 2001.

\bibitem{Fernandez_de_la_Vega:2003}
W.~Fernandez de~la Vega, Marek Karpinski, Claire Kenyon, and Yuval Rabani.
\newblock Approximation schemes for clustering problems.
\newblock In {\em Proceedings of the thirty-fifth annual ACM Symposium on
  Theory of computing, STOC}, pages 50--58. ACM Press, 2003.

\bibitem{Fiedler:1973}
M.~Fiedler.
\newblock Algebraic connectivity of graphs.
\newblock {\em Czechoslovak Math. J.}, 23:298--305, 1973.

\bibitem{Flake:2002}
G.~W. Flake, S.~Lawrence, C.~L. Giles, and F.~M. Coetzee.
\newblock Self-organization and identification of web communities.
\newblock {\em Computer}, 35(3):66--71, 2002.

\bibitem{Fortunato:2004}
S.~Fortunato, V.~Latora, and M.~Marchiori.
\newblock {A Method to Find Community Structures Based on Information
  Centrality}.
\newblock {\em ArXiv:cond-mat/0402522}, February 2004.

\bibitem{Gaveau:1999}
B.~Gaveau, A.~Lesne, and L.~S. Schulman.
\newblock Spectral signatures of hierarchical relaxation.
\newblock {\em Physics Letters A}, 258(4-6):222--228, July 1999.

\bibitem{Girvan_Newman:2002}
M.~Girvan and M.~E.~J. Newman.
\newblock Community structure in social and biological networks.
\newblock {\em PNAS}, 99(12):7821--7826, 2002.

\bibitem{HCI03}
Micka\"el Hoerdt and Damien Magoni.
\newblock Completeness of the internet core topology collected by a fast
  mapping software.
\newblock In {\em Proceedings of the 11th International Conference on Software,
  Telecommunications and Computer Networks}, pages 257--261, Split, Croatia,
  October 2003.

\bibitem{Jambu}
M~Jambu and Lebeaux M.-O.
\newblock {\em Cluster analysis and data analysis}.
\newblock North Holland Publishing, 1983.

\bibitem{Kernighan:1970}
B.~W. Kernighan and S.~Lin.
\newblock An efficient heuristic procedure for partitioning graphs.
\newblock {\em Bell System Technical Journal}, 49(2):291--308, 1970.

\bibitem{kleinberg:2001}
Jon Kleinberg and Steve Lawrence.
\newblock The structure of the web.
\newblock {\em Science}, 294(5548):1849--1850, 2001.

\bibitem{Lovasz_random_walks}
L.~Lov{\'a}sz.
\newblock Random walks on graphs: a survey.
\newblock In {\em Combinatorics, Paul Erd\H os is eighty, Vol.\ 2 (Keszthely,
  1993)}, volume~2 of {\em Bolyai Soc. Math. Stud.}, pages 353--397. J\'anos
  Bolyai Math. Soc., Budapest, 1996.

\bibitem{Newman:2003}
M.~E.~J. Newman.
\newblock The structure and function of complex networks.
\newblock {\em SIAM REVIEW}, 45:167, 2003.

\bibitem{Newman:2004}
M.~E.~J. Newman.
\newblock Fast algorithm for detecting community structure in networks.
\newblock {\em Physical Review E (Statistical, Nonlinear, and Soft Matter
  Physics)}, 69(6):066133, 2004.

\bibitem{Newman_Girvan:2004}
M.~E.~J. Newman and M.~Girvan.
\newblock Finding and evaluating community structure in networks.
\newblock {\em Physical Review E (Statistical, Nonlinear, and Soft Matter
  Physics)}, 69(2):026113, 2004.

\bibitem{Pothen:1990}
A.~Pothen, H.~D. Simon, and K.-P. Liou.
\newblock Partitioning sparse matrices with eigenvectors of graphs.
\newblock {\em SIAM J. Matrix Anal. Appl.}, 11(3):430--452, 1990.

\bibitem{Radicchi_Filippo:2004}
F.~Radicchi, C.~Castellano, F.~Cecconi, V.~Loreto, and D.~Parisi.
\newblock {Defining and identifying communities in networks}.
\newblock {\em PNAS}, 101(9):2658--2663, 2004.

\bibitem{Ravasz:2002}
E.~Ravasz, A.~L. Somera, D.~A. Mongru, Z.~N. Oltvai, and A.-L. Barab\'asi.
\newblock {Hierarchical Organization of Modularity in Metabolic Networks}.
\newblock {\em Science}, 297(5586):1551--1555, 2002.

\bibitem{Schulman:2001}
L.~S. Schulman and B.~Gaveau.
\newblock Coarse grains: The emergence of space and order.
\newblock {\em Foundations of Physics}, 31(4):713--731, April 2001.

\bibitem{Simonsen:2004}
I.~Simonsen, K.~Astrup Eriksen, S.~Maslov, and K.~Sneppen.
\newblock Diffusion on complex networks: a way to probe their large-scale
  topological structures.
\newblock {\em Physica A: Statistical Mechanics and its Applications},
  336(1-2):163--173, May 2004.

\bibitem{Strogatz:2001}
S.~H. Strogatz.
\newblock Exploring complex networks.
\newblock {\em Nature}, 410:268--276, March 2001.

\bibitem{Ward:1963}
J.~H. Ward.
\newblock Hierarchical grouping to optimize an objective function.
\newblock {\em Journal of the American Statistical Association},
  58(301):236--244, 1963.

\bibitem{wasserman94socialnetwork}
S.~Wasserman and K.~Faust.
\newblock {\em Social network analysis}.
\newblock Cambridge University Press, Cambridge, 1994.

\bibitem{url_prog}
\verb#http://liafa.jussieu.fr/~pons/#.

\end{thebibliography}
\bibliographystyle{plain}

\newpage
\appendix 
\section{Proofs of previously known results}
\paragraph{Proof of Property \ref{limit_proba}\\}
$$
P = \sum_{\alpha = 1}^n \lambda_{\alpha}v_{\alpha}u_{\alpha}^T \textrm{, and } P^t = \sum_{\alpha = 1}^n \lambda_{\alpha}^tv_{\alpha}u_{\alpha}^T \textrm{, and so } P_{ij}^t = \sum_{\alpha = 1}^n \lambda_{\alpha}^tv_{\alpha}(i)u_{\alpha}(j)
$$
When $t$ tends towards infinity, all the terms $\alpha \geq 2$ vanish. It is easy to show that the first right eigenvector $v_1$ is constant. By normalizing we have $\forall i, v_1(i) = \frac{1}{\sqrt{\sum_k d(k)}}$ and $\forall j, u_1(j) = \frac{d(j)}{\sqrt{\sum_k d(k)}}$. We obtain:
$$
\lim_{t \rightarrow +\infty} P_{ij}^t = \lim_{t \rightarrow +\infty} \sum_{\alpha = 1}^n \lambda_{\alpha}^t v_{\alpha}(i) u_{\alpha}(j) = v_1(i) u_1(j) = \frac{d(j)}{\sum_{k=1}^nd(k)}
$$

\paragraph{Proof of Property \ref{symetry_proba}\\}
This property can be written as the matricial equation $DP^tD^{-1} = (P^t)^T$ (where $M^T$ is the transpose of the matrix $M$). By using $P = D^{-1}A$ and the symmetry of the matrices $D$ and $A$, we have: $DP^tD^{-1} = D(D^{-1}A)^tD^{-1} = (AD^{-1})^t = (A^T(D^{-1})^T)^t = ((D^{-1}A)^T)^t = (P^t)^T$.

\paragraph{Proof of Theorem \ref{th_delta_sigma_1}\\}
First notice that the distance $r$ can be considered as a metric in $\mathbb{R}^n$ (that contains the probability vectors $P_{C\point}$) associated to an inner product $<.|.>$. We have: $r_{iC}^2 = <P_{C\point}^t - P_{i\point}^t|P_{C\point}^t - P_{i\point}^t>$.  In order to clarify the text we will use vectorial notation. For all vertex $i$ and community $C$, we define $\overrightarrow{iC} = P_{C\point}^t - P_{i\point}^t$ and for any two communities $C_1$ and $C_2$, $\overrightarrow{C_1C_2} = P_{C_2\point}^t - P_{C_1\point}^t$. We can write:
$$
\sum_{i\in C_1} r_{iC_3}^2 =  \sum_{i\in C_1} < \overrightarrow{iC_3} | \overrightarrow{iC_3} > = \sum_{i\in C_1} \Big( <\overrightarrow{iC_1}|\overrightarrow{iC_1}> + 2 <\overrightarrow{iC_1}|\overrightarrow{C_1C_3}> + <\overrightarrow{C_1C_3}|\overrightarrow{C_1C_3}> \Big)
$$
We then notice that $P_{C_1\point}^t$ is the centroid of the vectors $\{P_{i\point}^t | i \in C_1\}$, therefore we have $\sum_{i\in C_1} \overrightarrow{iC_1} = \overrightarrow 0$. Moreover we also have $\overrightarrow{C_1C_3} = \frac{|C_2|}{|C_1| + |C_2|} \overrightarrow{C_1C_2}$ and we finally obtain:
$$
\sum_{i\in C_1} r_{iC_3}^2 = \sum_{i\in C_1} r_{iC_1}^2 + \frac{|C_1||C_2|^2}{(|C_1| + |C_2|)^2} r_{C_1C_2}^2
$$
This also holds if we replace $C_1$ by $C_2$ and $C_2$ by $C_1$. Therefore:
$$
\sum_{i\in C_3} r_{iC_3}^2 = \sum_{i\in C_1} r_{iC_3}^2 + \sum_{i\in C_2} r_{iC_3}^2 = \sum_{i\in C_1} r_{iC_1}^2 + \sum_{i\in C_2} r_{iC_2}^2 + \frac{|C_1||C_2|}{|C_1| + |C_2|} r_{C_1C_2}^2
$$
We deduce the claim by replacing this expression into Equation (\ref{def_delta_sigma}).

\paragraph{Proof of Theorem \ref{th_delta_sigma_2}\\}
We replace the four $\Delta\sigma$ of Equation~(\ref{eq_th_delta_sigma_2}) by their values given by Theorem~\ref{th_delta_sigma_1}. We multiply each side by $\frac{n(|C_1|+|C_2|+|C|)}{|C|}$ and use $|C_3| = |C_1|+|C_2|$, and obtain the equivalent equation:
$$
(|C_1|+|C_2|)r_{C_3C}^2 = |C_1|r_{C_1C}^2 + |C_2|r_{C_2C}^2 - \frac{|C_1||C_2|}{|C_1|+|C_2|}r_{C_1C_2}^2
$$
Then we use the fact that $P_{C_3\point}^t$ is the barycenter of $P_{C_1\point}^t$ weighted by $|C_1|$ and of $P_{C_2\point}^t$ weighted by $|C_2|$, therefore: 
$$
|C_1|r_{C_1C}^2 + |C_2|r_{C_2C}^2 = (|C_1|+|C_2|)r_{C_3C}^2 + |C_1|r_{C_1C_3}^2 + |C_2|r_{C_2C_3}^2
$$
We conclude using $|C_1|r_{C_1C_3}^2 + |C_2|r_{C_2C_3}^2 = \frac{|C_1||C_2|}{|C_1|+|C_2|}r_{C_1C_2}^2$.

\section{More experimental results}
\begin{figure}[ht]
\begin{center}
\includegraphics{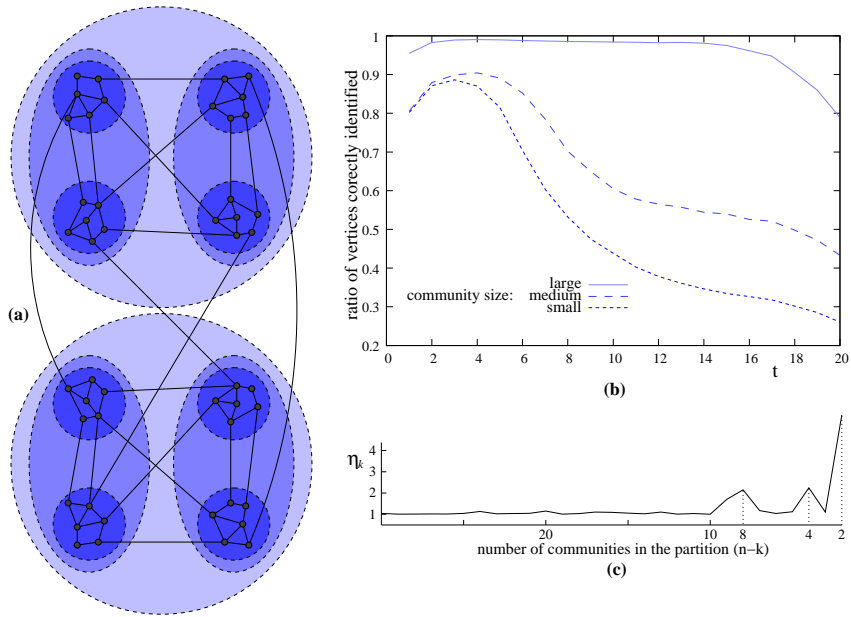}
\caption{(a) Hierarchical community structure used for the test. Each of the $8$ small communities has 32 vertices. (b) The three ratios (corresponding to the three community sizes) of vertices correctly identified by our algorithm as a function of the random walks length $t$ used. (c) Evolution of $\eta_k$ (last $30$ steps) showing that we identify the three scales in the community structure.}
\label{figure:hierarchical}
\end{center}
\end{figure}

\paragraph{Influence of the length $t$ on a hierarchically structured network.}
In this appendix we study the influence of the length $t$ of the random walks. To do this we need graphs with a hierarchical community structure. We generated graphs with $n = 256$ vertices divided into $8$ small communities included in $4$ medium communities and then in $2$ large communities, see Figure~\ref{figure:hierarchical}(a). We chose an internal edge density in small communities $p_1 = 0.3$, and an edge density between small, medium and large communities $p_2 = 0.1, p_3 = 0.75$ and $p_4 = 0.5$, respectively. We ran our algorithm for different $t$ and we computed the ratio of vertices correctly identified for the three kinds of communities. The results (Figure~\ref{figure:hierarchical}(b)) show that we get good performances for short random walks. We also notice that the range of $t$ that gives good results depends on the size of the communities found. There is a relation between the size of the communities to identify and the value of $t$ that we must choose. It seems that a good choice of $t$ must leave enough time to random walks to reach all the vertices of a community but not enough time to reach all the vertices of the graph. This is why a length close to the diameter of the communities to identify seems a relevant choice. Moreover these tests show that our approach is able to identify community structures at different scales: we clearly have three peaks on $\eta_k$ corresponding to the three sizes of communities (Figure~\ref{figure:hierarchical}(c)).

\paragraph{Experiment on a real network (Internet map).} We tested our algorithm on a map of the Internet (provided by Magoni \cite{HCI03}) that contains 12,929 routers and 52,844 physical links between them. Each router belongs to a known Autonomous system (AS) and the aim of the experiment is to see if we can retrieve them using our community detection algorithm, exploring the idea that they may correspond to dense subgraphs.
 
The map has been established from intensive {\em traceroute} experiments and only covers a part of the Internet: routers from 383 different AS are represented and are linked by 35,096 internal links and 17,748 links between different AS. However, due to the measurement method, some AS are poorly discovered: we only see a small number of their routers and their internal links. This phenomenon implies that many small AS (small on the map but not necessarily in reality) cannot be considered as communities (from our point of view). For instance, 320 of the represented AS (corresponding to 5,186 routers) have the ratio of number of external edges by the number of internal edges larger than 1 and 237 AS (2,706 routers) have this ratio larger than 2.

Our algorithm computed a community structure for $t = 5$ in 5 minutes (on a P4-M 2.2\ Ghz, 512\ MB). We looked at the partition with the best modularity ($Q = 0.73$), which contains 646 communities. For each router, we computed the ratio of routers in its community that belong to the same AS. The mean of this ratio over the routers is 52\%, which shows that even in these bad conditions our algorithm is able to group together a significant portion of the router of an AS.

\end{document}